\documentclass[conference]{IEEEtran}
\IEEEoverridecommandlockouts
\PassOptionsToPackage{bookmarks=false}{hyperref}
\renewcommand{\figurename}{Figure}

\ifCLASSINFOpdf
  \usepackage[pdftex]{graphicx}
\else

\fi

\usepackage[latin1,utf8]{inputenx}
\usepackage{enumitem}
\usepackage[cmex10]{amsmath}
\usepackage[ruled,vlined]{algorithm2e}

\SetCommentSty{mycommfont}
\usepackage{array}
\usepackage{xcolor}
\usepackage[colorlinks=true,citecolor=blue,linkcolor=blue,urlcolor=blue,plainpages=false]{hyperref}
\usepackage{url}
\usepackage{booktabs, tabularx, multirow, array}
\usepackage{breakurl}
\usepackage{nomencl}
\usepackage{multirow}
\usepackage{comment} 
\usepackage{amsmath}
\usepackage{amsfonts}
\usepackage{amssymb}
\usepackage{graphicx}
\usepackage{amsthm}
\usepackage{eurosym}
\usepackage{cite}
\usepackage{siunitx}
\usepackage{float}
\usepackage{subcaption}
\floatstyle{ruled}
\newfloat{model}{thp}{lop}
\floatname{model}{\textsc{Model}}
\usepackage{bm}
\DeclareMathOperator*{\argmax}{arg\,max}
\DeclareMathOperator*{\argmin}{arg\,min}

\usepackage{tikz}
\usetikzlibrary{arrows.meta}
\usetikzlibrary{positioning}
\usetikzlibrary{graphs}
\tikzset{main/.style={circle,
                            draw,
                            minimum size=0.5 cm,
                            inner sep=0pt}}
\tikzset{dots/.style={circle,
                            minimum size=1 cm,
                            inner sep=0pt}}

\DeclareSIUnit{\million}{\text{million}}

\def\F{\bm{\mathcal{F}}}
\def\R{{\bm{R}}}

\usepackage{etoolbox}
\renewcommand\nomgroup[1]{%
  \item[\bfseries
  \ifstrequal{#1}{I}{\textit{Indices and Sets}}{%
  \ifstrequal{#1}{P}{\textit{Parameters}}{%
  \ifstrequal{#1}{F}{\textit{Functions}}{%
  \ifstrequal{#1}{M}{\textit{Production Models}}{%
  \ifstrequal{#1}{V}{\textit{Variables}
  }{}}}}}%
]}
\usepackage[capitalise]{cleveref}

\makenomenclature
\begin{document}

\IEEEoverridecommandlockouts

\title{QuickFlex: a Fast Algorithm for Flexible Region Construction for the TSO-DSO Coordination}

\author{
\IEEEauthorblockN{Luis Lopez, Alvaro Gonzalez-Castellanos, David Pozo}
\IEEEauthorblockA{Center for Energy Science and Technology \\
Skolkovo Institute of Science and Technology (\textit{Skoltech})}
\and
\IEEEauthorblockN{Mardavij Roozbehani, Munther Dahleh}
\IEEEauthorblockA{Institute for Data, Systems, and Society \\
Massachusetts Institute of Technology (\textit{MIT})}}
\maketitle

\begin{abstract}
Most of the new technological changes in power systems are expected to take place in distribution grids. The enormous potential for distribution flexibility could meet the transmission system's needs, changing the paradigm of generator-centric energy and ancillary services provided to a demand-centric one, by placing more importance on smaller resources, such as flexible demands and electric vehicles. 
For unlocking such capabilities, it is essential to understand the aggregated flexibility that can be harvested from the large population of new technologies located in distribution grids. Distribution grids, therefore, could provide aggregated flexibility at the transmission level.  
To date, most computational methods for estimating the aggregated flexibility at the interface between distribution grids and transmission grids have the drawback of requiring significant computational time, which hinders their applicability.
This paper presents a new algorithm, coined as \texttt{QuickFlex}, for constructing the flexibility domain of  distribution grids. Contrary to previous methods, a priory flexibility domain accuracy can be selected.
Our method requires few iterations for constructing the flexibility region. The number of iterations needed is mainly independent of the distribution grid's input size and flexible elements. Numerical experiments are performed in four grids ranging from 5 nodes to 123 nodes. It is shown that QuickFlex outperforms existing proposals in the literature in both speed and accuracy. 

\end{abstract}

\begin{IEEEkeywords}
QuickFlex, Flexibility Region, TSO/DSO interface, PQ region, Active Distribution Grids
\end{IEEEkeywords}
\printnomenclature[0.55in]

\IEEEpeerreviewmaketitle

\nomenclature[I]{$i, N$}{Index/set of nodes}
\nomenclature[I]{$r, R$}{Index/set of reference buses}
\nomenclature[I]{$(i,j), L$}{Index/set of branches}
\nomenclature[I]{$k, G$}{Index/set of generators}
\nomenclature[I]{$d, D$}{Index/set of power demand}
\nomenclature[P]{$S^{gl}_k, S^{gu}_k $}{Generator complex power limits $k \in G$}
\nomenclature[P]{$C_{1k}, C_{0k}$}{Generator cost components $k \in G$}
\nomenclature[P]{$V^l_i, V^u_i$}{Voltage limits $i \in N$}
\nomenclature[P]{$S^d_k$}{Load complex power consumption $d \in D$}
\nomenclature[P]{$Y_{ij}, Y^c_{ij}$}{Branch pi-section parameters $(i,j) \in L$}
\nomenclature[P]{$S^u_{ij} $}{Branch apparent power limit $(i,j) \in L$}

\nomenclature[V]{$\text{Variables are defined within the models.}$}{}

\section{Introduction}\label{Sec:Intro}
Distributed energy resources (DERs) connected to the distribution grid have seen a fast widespread in recent years, such as electric vehicles (EVs), inverted-based distributed generators (e.g., photovoltaic power plants and micro wind turbines), and heat pumps \cite{nosratabadi2017comprehensive}. 
However, on the one hand, transmission system operators (TSOs) have minor, or no knowledge about the capacity, type, characteristics, and generation and consumption patterns of the DERs connected to the distribution level\cite{neuhoff2018tso}. Lack of information leads to load, and generation forecasting errors, affecting the overall system operation \cite{mahmud2020internet}. In several power markets,  
distribution system operators (DSOs) are responsible for managing and collecting all information from DERs and are therefore able to assess their potential flexibility capabilities, which can be in turn be offered to the TSO \cite{capitanescu2018tso}.
Increased interaction between DSOs and TSOs improves system flexibility, reducing and delaying investments in network reinforcement \cite{silva2018estimating}.

\subsection{Literature Review}
An approach for the representation of the flexibility available at the DSO/TSO interface, that has gained interest in recent years, has been the use of \textit{flexible operating regions} \cite{riaz2019feasibility, gonzalez2018determination,AgeevaREEPE,ageeva2020coordination}.
The feasible operating region in a distribution network indicates the region where active and reactive power exchanges can be operated without violating the steady-state network's grid limits \cite{givisiez2020review}.
Information about the flexibility operation region at any time can be used to provide ancillary services. Thus, a flexible operation region would help balance the entire system by employing all connected resources more efficiently \cite{silva2018challenges}.

To date, different computation methods have been proposed to estimate the feasible operating region at the DSO/TSO interface \cite{contreras2021}. 
Methods based on random sampling (Monte Carlo),  point-wise sum of feasible regions (Minkowsky), and iterative optimization-based algorithms are the main methodological directions employed for the aggregation of operational flexibility in distribution networks. 
Some of the methodological gaps from these methods are: 
$\bm{(i)}$ \textit{Monte Carlo methods}  are unable to efficiently recover the feasibility region \cite{riaz2019feasibility,AgeevaREEPE}. Their accuracy depends on the size of the random input vector for the operation points, which grow rapidly with the number of connected DERs. Previous works have reported orders of millions of sampling points \cite{contreras2021}. 
$\bm{(ii)}$ Methods based on the \textit{Minkowsky addition} (point-wise) sum the feasible operation regions of individuals system's resources. They can be computationally demanding \cite{silva2018challenges}.
Besides, operational constraints, such as line and voltage limits, are not generally considered, so there is no accurate representation of the flexibility available that recognizes power flow constraints and technical grid limits resulting in overoptimistic flexible regions.
$\bm{(iii)}$ Finally, \textit{iterative optimization-based methods} have gained popularity for building flexible regions \cite{ageeva2020coordination,capitanescu2018tso,pisciella2017optimal}.
These methods solve a sequence of optimization problems to recover the boundary of the flexible operating regions. They have embedded the operational grid limits and the power flow equations. These methods may have higher computational costs in the calculation of the flexibility region. 

\subsection{Paper Contributions and Organization}
This paper aims to discuss the coordination between TSO and DSO driven by the flexible resources located in distribution networks. Thus, this paper focuses on determining the feasible operation region of distribution networks with flexible elements that can be coordinated. 
We propose an algorithm for the fast calculation of the flexibility region at the TSO/DSO interface that considers distribution grid operational limits and Kirchoff's physic laws. The proposed algorithm is coined as \texttt{QuickFlex}. 
The QuickFlex provides an error bound at each iteration step on the accuracy of the constructed flexibility region. The main contributions of our work are listed as follows: 
\begin{enumerate}[label=(\roman*)]
    \item We propose the QuickFlex algorithm for computing the feasible region at the TSO/DSO interface. It generates feasible regions with few iterations and with measurable accuracy. Additionally, we empirically show that QuickFlex's iterations required are independent of the number of DERs connected to the network. 
    \item We evaluate the impact of using power flow relaxations and approximations in the computation of the feasibility. In particular, we considered using a DistFlow formulation, a second-order cone relaxation of the DistFlow, and linear approximation to the power flow equations. 
    \item We evaluate the proposed method on four test distribution networks over a set of three power grids. The QuickFlex algorithm is also compared against the Monte Carlo method \cite{riaz2019feasibility,AgeevaREEPE}, epsilon-constrained method \cite{ageeva2020coordination,capitanescu2018tso},  and radial reconstruction method \cite{pisciella2017optimal}.
\end{enumerate}

The remainder of the paper is organized as follows. \cref{Sec:Grid_model} corresponds to the modeling of the networks and the OPF. \cref{Sec:Methods} describes the methodology and the proposed algorithm. \cref{Sec:Case_Study} shows the case study. Finally, \cref{Sec:Conclusions} presents the conclusions. 

\section{Grid-aware Aggregator Modeling}\label{Sec:Grid_model}

\subsection{The AC Optimal Power Flow}
In its basic form, the optimal power flow (OPF) consists of dispatching a series of generators for satisfying electric load demands while considering constraints that model the power grid physic laws and operational limits imposed on the grid and dispatchable elements. 
The objective function in the OPF is typically an active power loss minimization, or generation cost minimization \cite{gan2014exact}.

\begin{model}[h]
\caption{AC Optimal Power Flow \hfill [NLP]}
\label{Mod:opf}

\mbox{\bf Variables: } 
\begin{IEEEeqnarray}{lll}
& s^g_k &\;\;k\in G \qquad
\mbox{ Generator complex power }
\nonumber
\label{var_generation}\\
& v_i &\;\;i\in N \qquad
\mbox{ Node complex voltage}
\nonumber
\label{var_voltage} \\
& s_{ij} &\;\;(i,j) \in L \;\;
\mbox{ Branch complex power flow}
\nonumber
\label{var_complex_power} 
\end{IEEEeqnarray}

\mbox{\bf Objective: }
\begin{IEEEeqnarray}{lll}
\text{min} \; & \sum_{k \in G} C_{1k}\Re(s^g_k) + C_{0k} & 
\IEEEyesnumber
\IEEEyessubnumber
\label{eq:objective}
\end{IEEEeqnarray}

\mbox{\bf Subject to: }
\begin{IEEEeqnarray}{llr}
& S^{gl}_k \leq s^g_k \leq S^{gu}_k & \;\; k \in G
\IEEEyessubnumber
\label{eq:gen_bounds}\\
& |s_{ij}| \leq S^u_{ij} & \;\; (i,j) \in L 
\IEEEyessubnumber
\label{eq:thermal_limit}\\
& |v_{r}| = 1 & \;\; r \in R 
\IEEEyessubnumber
\label{eq:ref_bus}\\
& V^l_i \leq |v_i| \leq V^u_i & \;\; i \in N 
\IEEEyessubnumber
\label{eq:voltage_bounds}\\
& s_{i}^{g} - s_{i}^{d}=\sum_{(i, j) \in L} s_{i j} & \;\; i \in N 
\IEEEyessubnumber
\label{eq:node_current}\\
& s_{i j} = \left( Y_{ij} + Y^c_{ij}\right)^*\left|v_{i}\right|^{2}-Y_{i j}^{*} v_{i} v_{j}^{*} & \;\; (i, j) \in L 
\IEEEyessubnumber
\label{eq:power_from}
\end{IEEEeqnarray}
\end{model}


Model \ref{Mod:opf} summarizes the optimal power flow formulation where \eqref{eq:objective} is a linear function for active power generation costs. Constraints \eqref{eq:gen_bounds} and \eqref{eq:thermal_limit} impose the generator output and line flow limits. Constraints \eqref{eq:ref_bus} and \eqref{eq:voltage_bounds} capture the voltage magnitude limits. Finally, constraints \eqref{eq:node_current} and \eqref{eq:power_from} capture the power flow's physical properties through Kirchoff's current laws and Ohm's Law. 
Note that the AC OPF is a non-linear and non-convex optimization problem.

\subsection{DistFlow. A specialized AC-OPF model for radial grids}

Typical distribution systems are centrally operated, ensuring a radial topology that provides power from a feeder to sequentially connected customers. 
The feeder or point of common coupling (PCC) interfaces with the transmission system.
A specialized AC-OPF for radial distribution grids is the so-called DistFlow formulation depicted in Model \ref{Mod: distflow}. It can be obtained by replacing the line flow equation \eqref{eq:power_from} with three expressions based on the distribution system's radial structure \cite{baran1989optimal}. First, the line losses are represented as \eqref{eq:losses_df}. Secondly, the voltage difference between nodes \eqref{eq:radial_flow_df} is set. Thirdly, the absolute square property of the apparent power is used to derive expression \eqref{eq:app_power_df}. Note that the DistFlow model is also a non-linear and non-convex optimization problem.

\begin{model}[h]
\caption{DistFlow \hfill[NLP]}
\label{Mod: distflow}

\mbox{\bf Variables: } $s^g_k,\; v_i, \; s_{ij}$
\begin{IEEEeqnarray}{lll}
& l_{ij} &\;\;(i,j) \in L \;\;
\mbox{ Current magnitude squared $|i_{ij}|^2$}
\nonumber
\label{var_complex_power_df} 
\end{IEEEeqnarray}

\mbox{\bf Objective: } \eqref{eq:objective}

\mbox{\bf Subject to: } \eqref{eq:gen_bounds}-\eqref{eq:node_current}
\begin{IEEEeqnarray}{llr}
& s_{i j} + s_{j i} = Z_{i j} l_{i j} & \;\; (i.j) \in L
\IEEEyesnumber
\IEEEyessubnumber
\label{eq:losses_df}\\
& v_i^2 - v_j^2 = \left(Z_{i j}^{*} s_{i j}+Z_{i j} s_{i j}^{*}\right)-\left|Z_{i j}\right|^{2} l_{i j} & (i, j) \in L
\IEEEyessubnumber
\label{eq:radial_flow_df}\\
&\left|s_{i j}\right|^{2} = |v_i|^2 l_{i j} & \; \; (i, j) \in L
\IEEEyessubnumber
\label{eq:app_power_df}
\end{IEEEeqnarray}
\end{model}

\subsection{A SOC DistFlow relaxation}

Convex relaxations of the power flow equations have brought a great deal of interest in recent years. These relaxations are attractive because they are computationally efficient and produce a feasible solution to the original non-convex problem. 
One of such relaxations is the second-order cone (SOC) relaxation of the DistFlow formulation \cite{jabr2006radial}.

The SOC power flow model is presented in Model \ref{Mod: SOC_DistFlow}. The nodal balance constraints, generator operating limits, and transmission line thermal limits are the same as in the DistFLow, while $w$ substitutes the product of the voltage variables. Therefore, constraint \eqref{eq:radial_flow_soc} captures the line power flow in the $w$-space and \eqref{eq:voltage_bounds_soc} sets the voltage constraints. Finally, constraint \eqref{eq:app_power_soc} enforces the relaxation of the line flow as a second-order cone constraint of the product of the square of the voltage and current. The resulting formulation is non-linear but convex. 

\begin{model}[h]
\caption{SOC DistFlow \hfill [SOCP]}
\label{Mod: SOC_DistFlow}

\mbox{\bf Variables: } $s^g_k,\; s_{ij}, \; l_{ij}$ \vspace{-0.5\jot}
\begin{IEEEeqnarray}{lll}
& w_i &\;\;i\in N \qquad
\mbox{ Node complex voltage}
\nonumber
\label{var_voltage}
\end{IEEEeqnarray}

\mbox{\bf Objective: } \eqref{eq:objective}

\mbox{\bf Subject to: } \eqref{eq:gen_bounds}-\eqref{eq:ref_bus},  \eqref{eq:node_current}, \eqref{eq:losses_df}
\begin{IEEEeqnarray}{llr}
& w_{i}-w_{j} = \left(Z_{i j}^{*} s_{i j}+Z_{i j} s_{i j}^{*}\right)-\left|Z_{i j}\right|^{2} l_{i j} & (i, j) \in L
\IEEEyessubnumber
\label{eq:radial_flow_soc}\\
& \left(V_{i}^{l}\right)^{2} \leq w_{i} \leq \left(V_{i}^{u}\right)^{2} & \;\; i \in N 
\IEEEyesnumber
\IEEEyessubnumber
\label{eq:voltage_bounds_soc}\\
&\left|s_{i j}\right|^{2} \leqslant w_{i} l_{i j} & \; \; (i, j) \in L
\IEEEyessubnumber
\label{eq:app_power_soc}
\end{IEEEeqnarray}
\end{model}

\subsection{LinDistFlow. A linear DistFlow approximation}

A linearized version of the DistFlow model, coined as LinDistFlow by the authors, was proposed in  \cite{baran1989optimal}. 
The LinDistFlow formulation, presented in Model  \ref{Mod: LinDistFlow}, assumes no losses and the branch voltage drops linearly dependant on the branch power flow.
Constraint \eqref{eq:thermal_limit_linf} decouples the components of the thermal line limits, while \eqref{eq:radial_flow_linf} is the reformulation of the branch voltage drop. Constraint \eqref{eq:losses_linf} represents the lack of power losses.

\begin{model}[h]
\caption{LinDistFlow \hfill [LP]}
\label{Mod: LinDistFlow}

\mbox{\bf Variables: }  $s^g_k,\; s_{ij}=p_{ij}+ \boldsymbol{j} q_{ij}, \; w_{i}$

\mbox{\bf Objective: } \eqref{eq:objective}

\mbox{\bf Subject to: } \eqref{eq:gen_bounds}, \eqref{eq:ref_bus}, \eqref{eq:node_current} \eqref{eq:voltage_bounds_soc}
\begin{IEEEeqnarray}{llr}
& \Re(s_{ij}) \leq p^u_{ij}, \; \; \Im(s_{ij}) \leq q^u_{ij} & \;\; (i,j) \in L 
\IEEEyesnumber
\IEEEyessubnumber
\label{eq:thermal_limit_linf}\\
& s_{i j}+s_{j i} = 0 & \;\; (i.j) \in L
\IEEEyessubnumber
\label{eq:losses_linf}\\
& w_{i}-w_{j} = \left(Z_{i j}^{*} s_{i j}+Z_{i j} s_{i j}^{*}\right)& (i, j) \in L
\IEEEyessubnumber
\label{eq:radial_flow_linf}
\end{IEEEeqnarray}
\end{model}

\section{Methodology}\label{Sec:Methods}
This section describes a new algorithm for the fast construction of TSO/DSO feasible regions. It is named \textit{QuickFlex} because of its similarity with the {\textit{QuickHull}}\footnote{\textit{QuickHull} algorithm is, at the same time, inspired in the \textit{Quicksort} algorithm for ordering vectors \cite{hoare1962quicksort}.} algorithm for building a convex hull from a set of data points \cite{barber1996quickhull}. 

\subsection{QuickFlex Algorithm}
We define $\F$ as the feasible operation region based on the AC-OPF formulation with no particular objective function, i.e., $\F = \{\eqref{eq:gen_bounds}-\eqref{eq:power_from}\}$. Alternatively, other OPF formulations, namely DistFlow, SOC DistFlow, and LinDistFlow, can be used in the same context for identifying the operational feasible set of a distribution grid.

Next, we define $\R$ as the two-dimensional PQ-space characterizing the aggregated flexibility at the TSO/DSO interface. The region $\R$ is a convex hull containing all feasible operating points of the distribution grid, $\R = \{(p_1,q_1) \mid (p_1,q_1) \in \F \} $, at TSO/DSO interface. Alternatively, we can represent $\R$ as the collection of operating points describing  the boundary on the convex hull, i.e.,  $\R = \{(p_1^\kappa,q_1^\kappa),   \mid (p_1^\kappa,q_1^\kappa) \in \F , \: \forall \kappa=1,\ldots, K\} $, where $K$ is the total number operational points forming its boundary. When not all the $K$ boundary operational points can be recovered, but only $k$ of them, we use $\R_k$ to denote an approximation of $\R$ with $k$ points.

\begin{figure}[h]
    \centering
    \includegraphics[width=0.9\columnwidth]{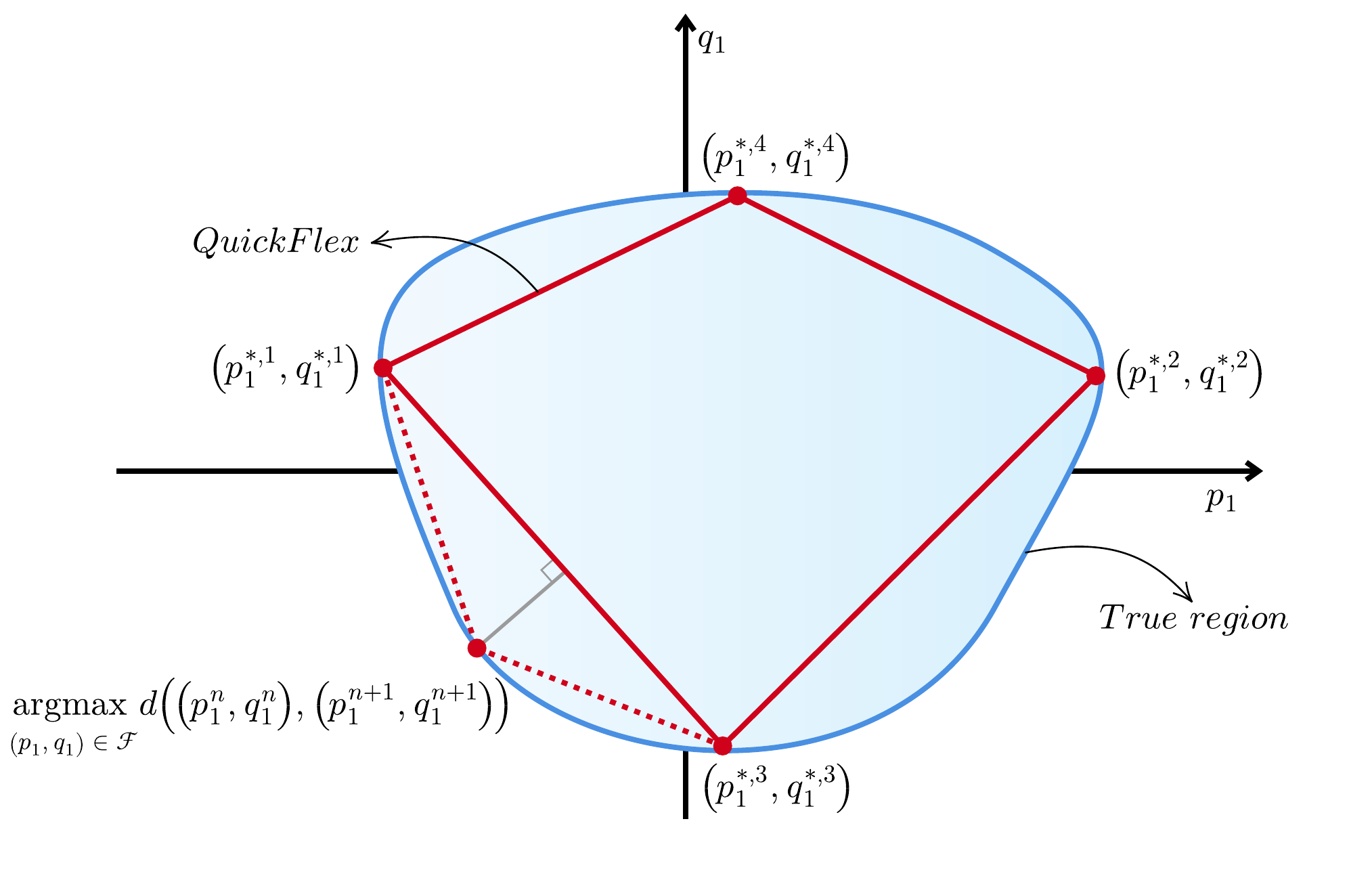}
    \caption{Illustrative  step-by-step QuickFlex execution for flexible region construction.}
    \label{fig:QH_Method}
\end{figure}

The idea of the QuickFlex algorithm is to iteratively find new points forming the boundary of the convex hull. The steps of the QuickFlex is illustrated in the Fig. \ref{fig:QH_Method} and summarized as follows:

\textbf{Step 0. Initialization.} Read data from grid and available flexibilities. Initialize feasible region $\R_0 = \emptyset$, and set the tolerance error $\epsilon$.

\textbf{Step 1. Initial convex hull of $\R$.} Find the minimum and maximum active and reactive power at the TSO/DSO interface that satisfy operational feasibility. Solutions of those four problems stated in \eqref{eq. initialization}, represent an initialization of the convex hull $\R$ formed by four vertices.
\begin{subequations} \label{eq. initialization}
\begin{IEEEeqnarray}{llr}
\R_1 \leftarrow  \R_0 \cup (p_1^{*1},q_1^{*1}) = \argmin_{(p_1,q_1)\in \F} \:  p_1  \\
\R_2 \leftarrow \R_1 \cup  (p_1^{*2},q_1^{*2}) = \argmax_{(p_1,q_1)\in \F}  \: p_1   \\
\R_3 \leftarrow \R_2 \cup (p_1^{*3},q_1^{*3}) = \argmin_{(p_1,q_1)\in \F}  \:  q_1  \\
\R_4 \leftarrow \R_3 \cup  (p_1^{*4},q_1^{*4}) = \argmax_{(p_1,q_1)\in \F}  \:  q_1  
\end{IEEEeqnarray}
\end{subequations}

\textbf{Step 2. Sequential update of $\R$}. For every facet of the current convex hull, visited in the clockwise direction, we find a new point that maximizes the perpendicular distance, $\bm{d}(\cdot)$, from that segment.  The new optimal point $(p_1^{*k},q_1^{*k})$ creates two new segments (facets) $(p_1^{*k},q_1^{*k})-(p_1^n,q_1^n)$ and $(p_1^{*k},q_1^{*k})-(p_1^{n+1},q_1^{n+1})$.
\begin{IEEEeqnarray}{llr} \label{eq. R_k}
\R_k \leftarrow \R_{k-1} \cup (p_1^{*k},q_1^{*k}), \:  \text{  where:  }  \nonumber  \\
(p_1^{*k},q_1^{*k}) 
= \argmax_{(p_1,q_1)\in \F} \:  \bm{d}\bm{\Big(}(p_1^n,q_1^n),(p_1^{n+1},q_1^{n+1}) \bm{\Big)} \:  
\end{IEEEeqnarray}

The idea of selecting the furthest point from the segment formed by the vertices $(p_1^n,q_1^n)$ and $(p_1^{n+1},q_1^{n+1})$ of the convex hull is to maximize the coverage of the selected feasible operating points.  

\textbf{Step 3. Stopping criteria and segment search elimination.} 
At every iteration, we calculate the area increase, $\varepsilon_k^A$, on the difference between the areas of the new convex hull, $A^{k}$ and the previous one $A^{k-1}$, \eqref{eq. stop}.
\begin{IEEEeqnarray}{llr} \label{eq. stop}
\varepsilon_k^A = \frac{A^{k} - A^{k-1}}{A^{k}} 
\end{IEEEeqnarray}
The area increase $\varepsilon_k^A$, is used for segment elimination and stopping criteria. 
If a segment does not contribute to increase the feasible area above a tolerance, \mbox{$\varepsilon_k^A \leq \epsilon$}, then it is excluded for further search. We repeat \textbf{Step 2} for the next segment and update the iteration counter $k \leftarrow k+1$.
Eventually, after eliminating all segments for searching new points, the feasible region is formed.  

%
\textbf{Output.} The result of this process is a set of points up to iteration $k$ forming the feasibility region $\R_k$.

Note that the QuickFlex is always selecting points from the boundary of the convex hull. No point is selected belonging to the interior or exterior of the convex hull. Contrary to Monte Carlo-based methods \cite{riaz2019feasibility}, and iterative optimization-based methods \cite{ageeva2020coordination,capitanescu2018tso}, QuickFlex does not require an a priori number of operational points for its evaluation, but rather a tolerance error on the TSO/DSO area. 

\section{Numerical Analysis}\label{Sec:Case_Study}
In this section, we analyze the proposed QuickFlex algorithm. 
Its performance is considered in several grids. We use the IEEE 5-node, IEEE 13-node, IEEE 37-node, and IEEE 123-node distribution test networks \cite{schneider2017analytic} to construct the feasibility regions at the TSO/DSO interface. 
For these networks, we have made two modifications. First, we have taken the balanced equivalents of these networks. Second, the network's flexible elements (DGs) are connected at the farthest nodes from the PCC. In the IEEE 13-node network, we have added four flexibility elements with an aggregated capacity of $150\%$ of the total installed demand. In the IEEE 37-node network, we have added six flexibility elements with a total installed capacity of twice the system demand. In the IEEE 123-node network, we have added $15$ flexibility elements with a total installed capacity of twice the demand. 

The computational experiments were implemented\footnote{ Source code is available at \href{https://github.com/Skoltech-PACO}{github.com/Skoltech-PACO}} in Julia with JuMP and executed on an Intel(R) Xeon(R) Gold 6148, CPU 2.40GHz, 2394 Mhz, 20 Core(s), 256 GB of RAM.

\subsection{Illustrative IEEE 5-node System}
To illustrate the QuickFlex algorithm, we first describe the algorithm in the IEEE 5-node test network \cite{sharma2017loss}. 

\begin{figure}[htbp]
    \centering
    \includegraphics[width=0.9\columnwidth]{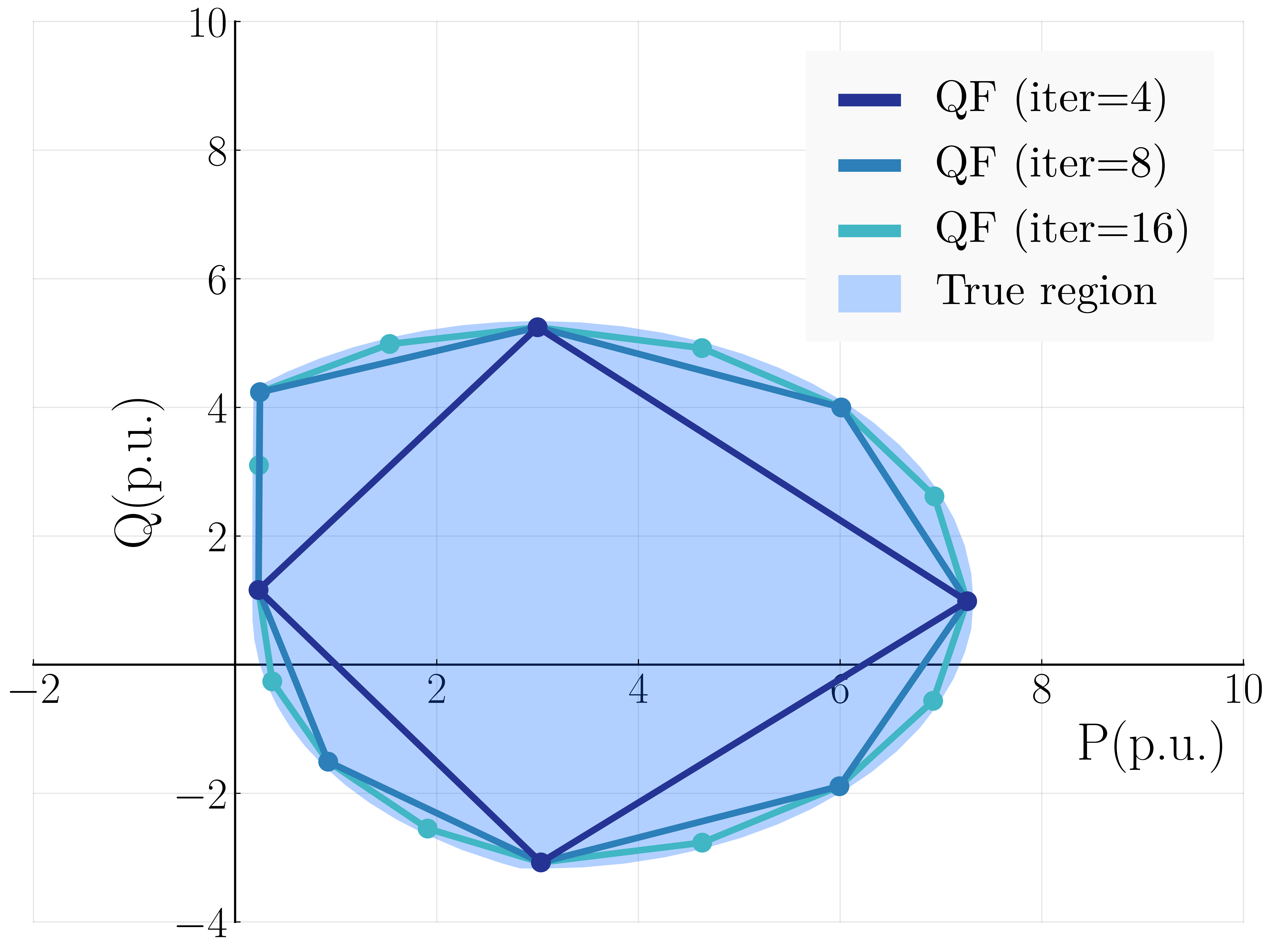}
    \caption{Flexible regions for the IEEE 5-node case study.}
    \label{fig:QF_Evol}
\end{figure}

\cref{fig:QF_Evol} shows the result of applying the QuickFlex algorithm. We have plotted on the same frame the state of the constructed region at different iterations. In iteration four, we have information on the maximum and minimum limits of active and reactive power supplied by the PCC. In iteration 16, we have a good estimate of the distribution system's feasibility region. 
The distribution system's true region is a reference polygon when 100 points have been obtained in QuickFlex. 

\subsection{Comparative Analysis on Power Flow Formulations}

\begin{figure*}[bht]
\centering
    \begin{minipage}{0.32\textwidth}
		\centering
    	\includegraphics[width=\textwidth]{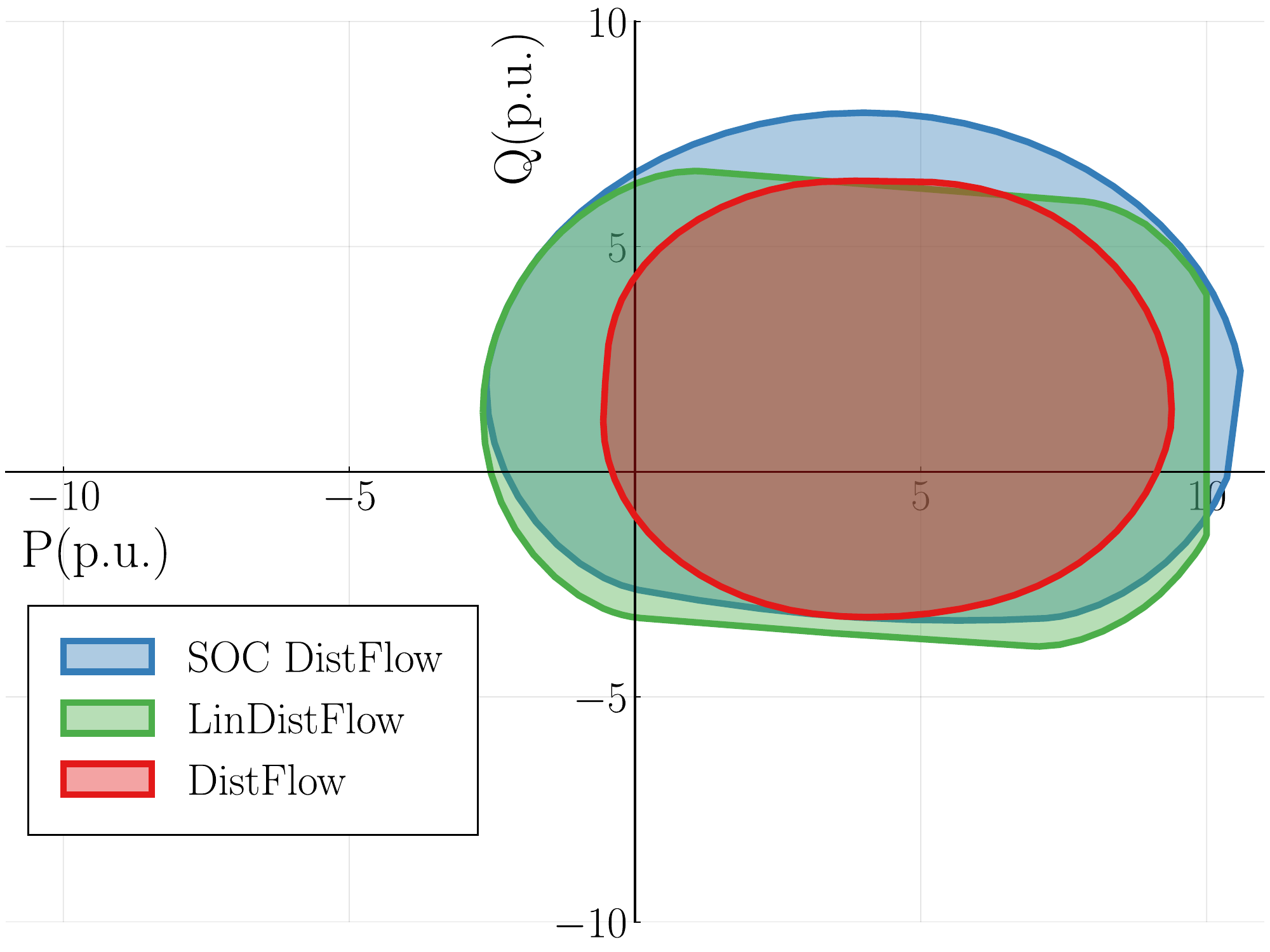}
	\end{minipage}
	\hspace{0.05cm}
	\begin{minipage}{0.32\textwidth}
	\includegraphics[width=\textwidth]{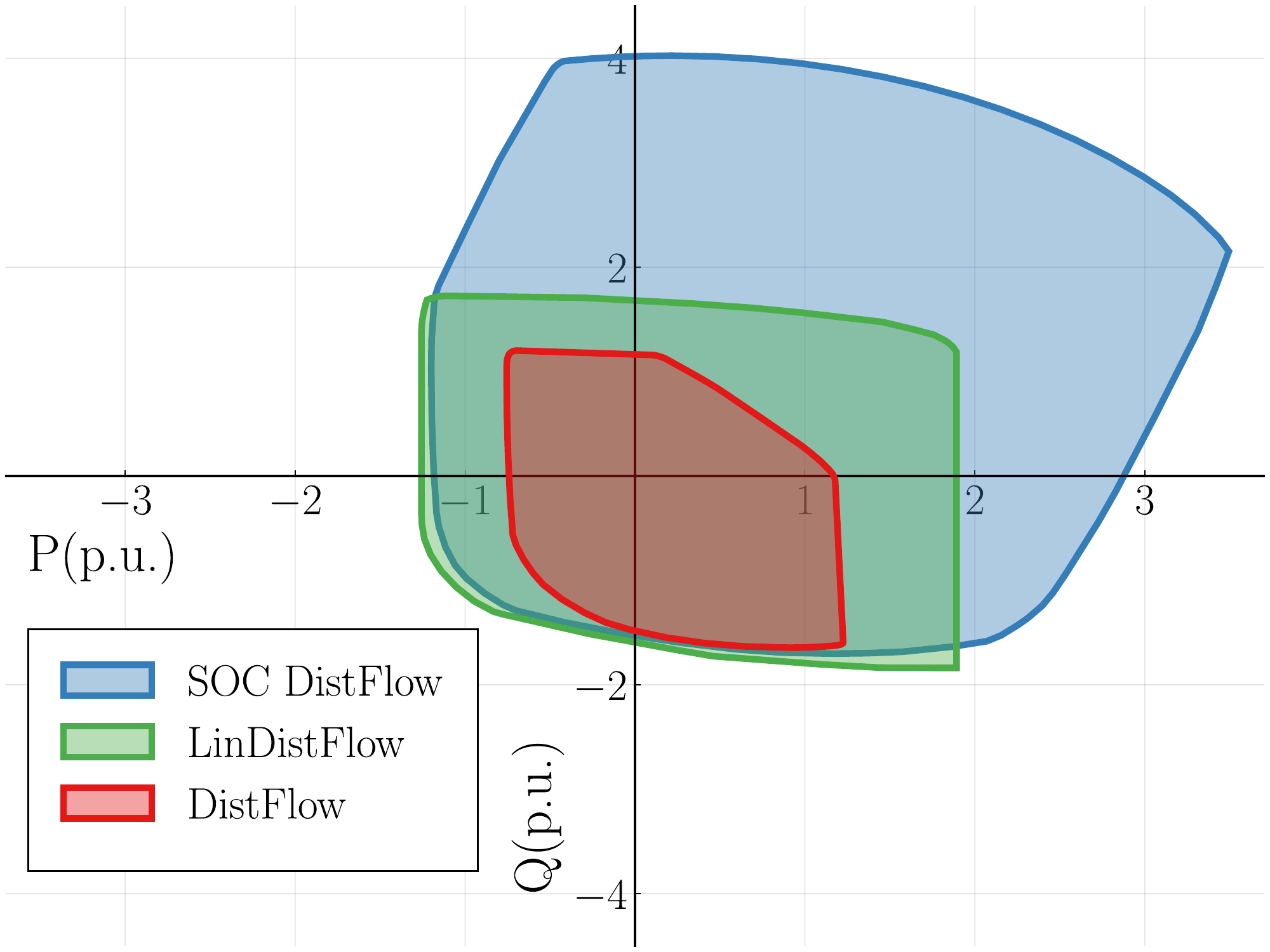}
	\end{minipage}
	\hspace{0.05cm}
	\begin{minipage}{0.32\textwidth}
		 \includegraphics[width=\textwidth]{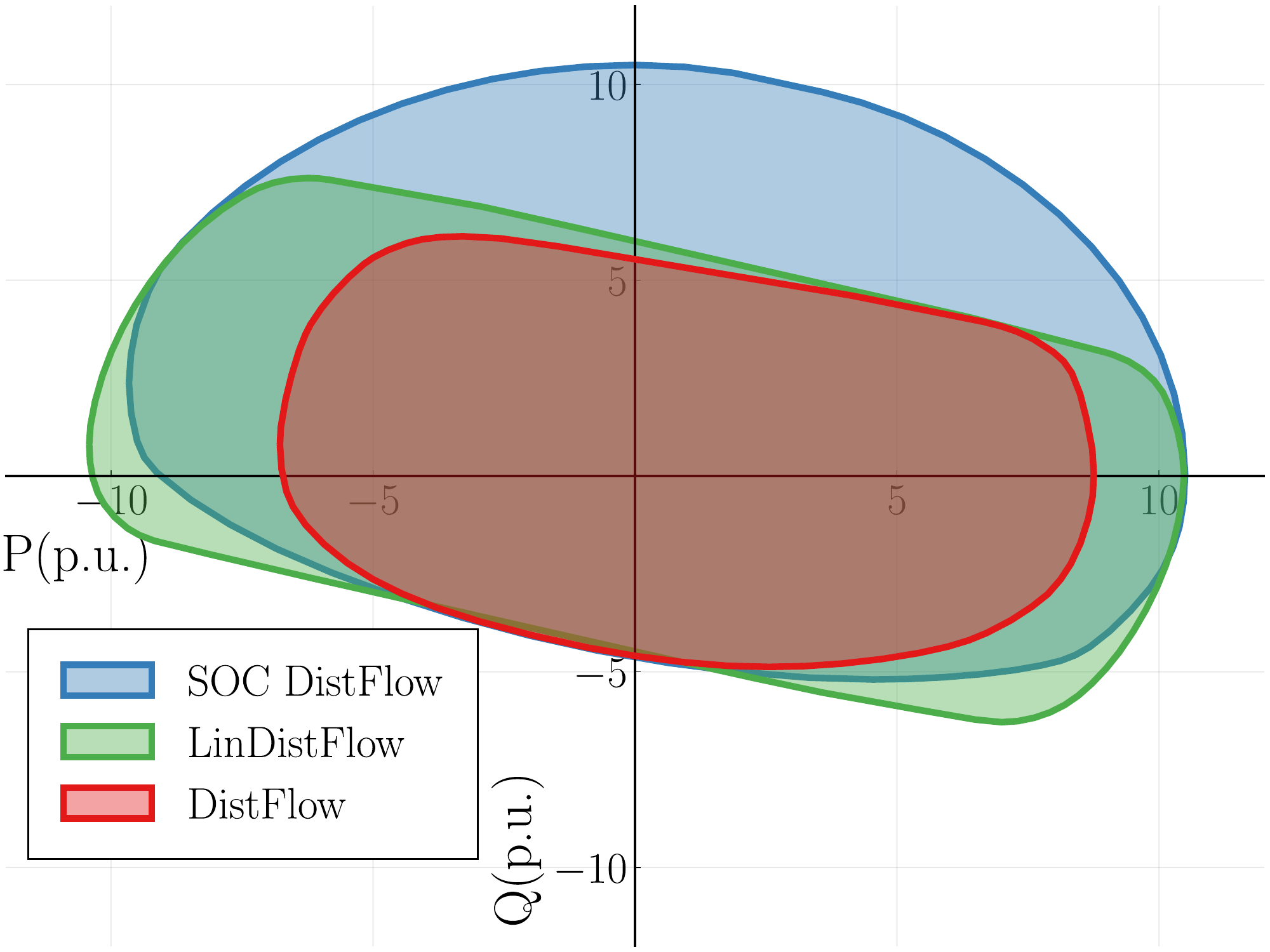}
	\end{minipage}
	\caption{Feasible regions at the TSO/DSO interface with different power flow formulations and distribution networks: (left) IEEE 13-node, (center) IEEE 37-node, and (right) IEEE 123-node systems.} 
	\vspace{-0.5cm}
	\label{fig:two-node-example}
\end{figure*}

For each of the IEEE 13-node, IEEE 37-node, and IEEE 123-node systems, we have evaluated the QuickFlex algorithm using the DistFlow formulation (exact solution), the SOC DistFlow formulation (convexified solution), and the LinDistFlow formulation (linearized solution). 
We have evaluated the feasibility region with a tolerance error of $\varepsilon = 10^{-3}$ for iteration area change.
\cref{fig:two-node-example} shows the feasibility regions obtained in each test system and the power flow models employed. We can observe that the relaxations/approximations overestimate the flexibility region at the PCC for the three distribution networks.

The QuickFlex algorithm's performance is summarized in  \cref{tab:Performance}. The first column, $k$  represents the total number of points computed. Each point has required to solve an optimization problem that considers feasible operational constraints $\F$. The second column, $\varepsilon_k^A$, is the area increase measured between the region computed in step $k-1$ and $k$ as a difference of areas. The third column provides the total time taken to solve the whole feasibility region. Finally, the fourth column, \textit{{Rel. Err. Area (\%)}}, summarizes the relative area error between different power flow formulations using the DistFlow as a reference. 

\begin{table}[h!]
  \caption{QuickFlex performance on three distribution grids and power flow formulations}
\centering
\vspace{0.1cm}
\begin{tabular}{llccc}
\cmidrule{2-5}  
\multicolumn{1}{l}{\textbf{IEEE 13-node}}  & $k$ & $\varepsilon_k^A$ (pu) & time (s) & {Rel. Err. Area  (\%)}\\
\midrule
 \textit{DistFlow} & 28    & 0.35$\cdot10^{-3}$ & 27.6 & -  \\
 \textit{SOC DistFlow } & 28    & 0.91$\cdot10^{-3}$ & 20.3 &   153.1\\
 \textit{LinDistFlow} & 29  & 0.22$\cdot10^{-3}$ & 17.5  & 146.1\\
\bottomrule
\vspace{0.1cm}
\end{tabular}
\begin{tabular}{llccc}
\cmidrule{2-5}  
\multicolumn{1}{l}{\textbf{IEEE 37-node}}  & $k$ & $\varepsilon_k^A$ (pu) & time (s) & {Rel. Err. Area  (\%)}\\
\midrule
\textit{DistFlow} & 18    & 0.49$\cdot10^{-3}$ & 155.1  & -  \\
\textit{SOC DistFlow} & 19    & 0.11$\cdot10^{-3}$ & 54.5  & 470.1\\
\textit{LinDistFlow} & 19    & 0.13$\cdot10^{-3}$ & 49.6 & 204.9 \\
\bottomrule
\vspace{0.1cm}
\end{tabular}
\begin{tabular}{llccc}
\cmidrule{2-5}  
\multicolumn{1}{l}{\textbf{IEEE 123-node}}  & $k$ & $\varepsilon_k^A$ (pu) & time (s) & {Rel. Err. Area  (\%)} \\
\midrule
\textit{DistFlow} & 23   & 0.37$\cdot10^{-3}$ & 292.2 & -  \\
\textit{SOC DistFlow} & 25   & 0.68$\cdot10^{-3}$ & 126.7  & 181.3\\
\textit{LinDistFlow} & 27   & 0.44$\cdot10^{-3}$ & 93.2  & 150.9\\
\bottomrule
\end{tabular}
\label{tab:Performance}    \vspace{-0.1cm}
\end{table}

As expected, computation time is lower when using the SOC DistFlow and LinDistFlow formulations. However, in both cases, the feasibility region is overestimated. It is worth highlighting that for the IEEE 37-node case, the convexified SOC DistFlow and the LinDistFlow formulations, respectively, provide almost five and two times flexible areas than the true one. 
The QuickFlex method does not depend on the size of the network neither does it depend on the number of flexible elements. The number of points ($k$) depends on the geometry of the region. 
A natural continuation of this research work is to investigate alternative formulations of power flow linearization for distribution networks \cite{bolognani2015fast,huang2021generalized}. We leave this analysis for future work.

\begin{figure}[h]
    \centering
    \includegraphics[width=0.8\columnwidth]{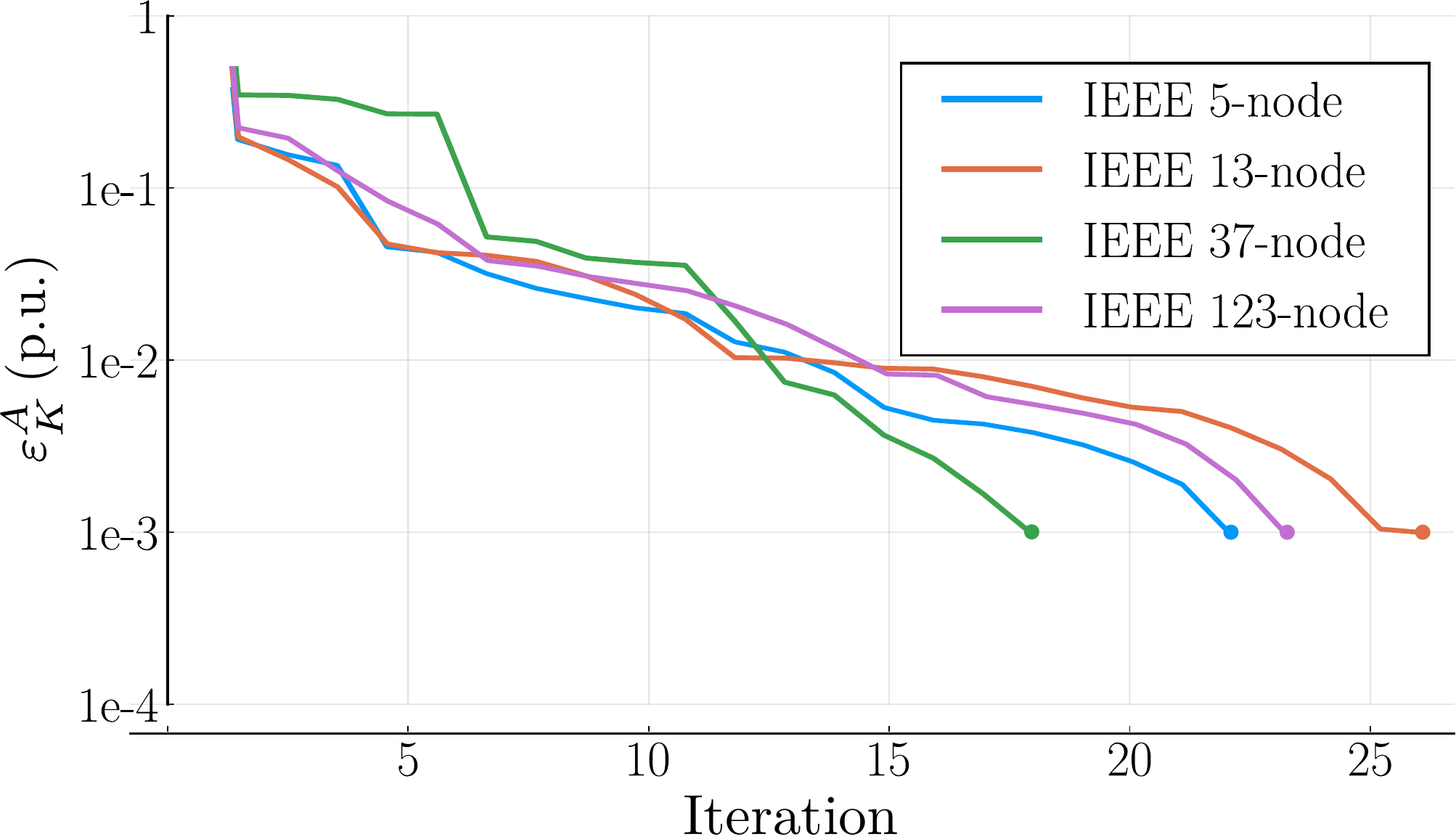}
    \caption{QuickFlex error evolution per iteration.}
    \label{fig:Tol_iter}
    \vspace{-0.5cm}
\end{figure}

 Finally, \cref{fig:Tol_iter} shows the error evolution of the feasibility region throughout the QuickFlex iterations. In the y-axis, in log-scale, it is represented feasible area increase between iterations. The iteration number is on the x-axis. 
We can see that the QuickFlex algorithm quickly reduces the area gains in the first few iterations. It is mainly independent of the size of the system as well as the number of flexible recourses.

\subsection{Comparative Analysis with Existing Methods}

In this section, we compare the QuickFlex algorithm against three other popular methods. 
In doing so, we first find the feasible TSO/DSO region with a given tolerance. 
Three distribution networks are used, the IEEE 13-node, IEEE 37-node, and IEEE 123-node grids. The feasible region is constructed based on the exact DistFlow formulation \eqref{Mod: distflow} for all cases.  The methods used for comparison are: 

\begin{itemize}
    \item[\textbf{QF}] \textit{QuickFlex Method.} It is solved with a tolerance error of $\epsilon = 10^{-3}$. The number of explored operating points $k$ is used as a reference in the other methods.  
    \item[\textbf{MC}] \textit{Monte Carlo Method} \cite{riaz2019feasibility,AgeevaREEPE}. In this case, we generate $k$ random samples from the flexible generators assuming a uniform distribution defined by their operational limits. 
    \item[\textbf{EC}] \textit{Epsilon-constrained Method}  \cite{ageeva2020coordination,capitanescu2018tso}. Based on the multiobjective optimization epsilon-constrained method, a sequence of maximization and minimization problems are solved where the active and reactive powers are fixed. The accuracy of the region depends on the granularity of the selected values. We divide the negative and positive axis in $\lceil{k/4}\rceil$ equidistant values. 
    \item[\textbf{RR}] \textit{Radial Reconstruction} \cite{pisciella2017optimal}. A sequence of problems is solved where the objective function is set to maximize a direction (fixed angle) in the PQ plane. The angle intervals between search directions are set by $360^{\circ}/k$.  
\end{itemize}
\noindent

\begin{table}[htbp]
  \centering
\caption{Percentage of flexible region recovered by the \textbf{MC}, \textbf{EC}, and \textbf{RR} methods} 
    \begin{tabular}{cccc}
\cmidrule{2-4}          & IEEE 13-node & IEEE 37-node & \multicolumn{1}{l}{IEEE 123-node} \\
    \midrule
    $k$ & 28    & 18    & 23 \\
    \midrule
    \textbf{MC} & 50.88 & 52.33 & 20.17 \\
    \textbf{EC} & 89.25 & 74.82 & 77.03 \\
    \textbf{RR} & 96.53 & 87.97 & 91.25 \\
    \bottomrule
    \end{tabular}%
  \label{tab:comp}%
\end{table}%

\begin{figure}
    \vspace{-0.5cm}
    \centering
    \includegraphics[width=0.8\columnwidth]{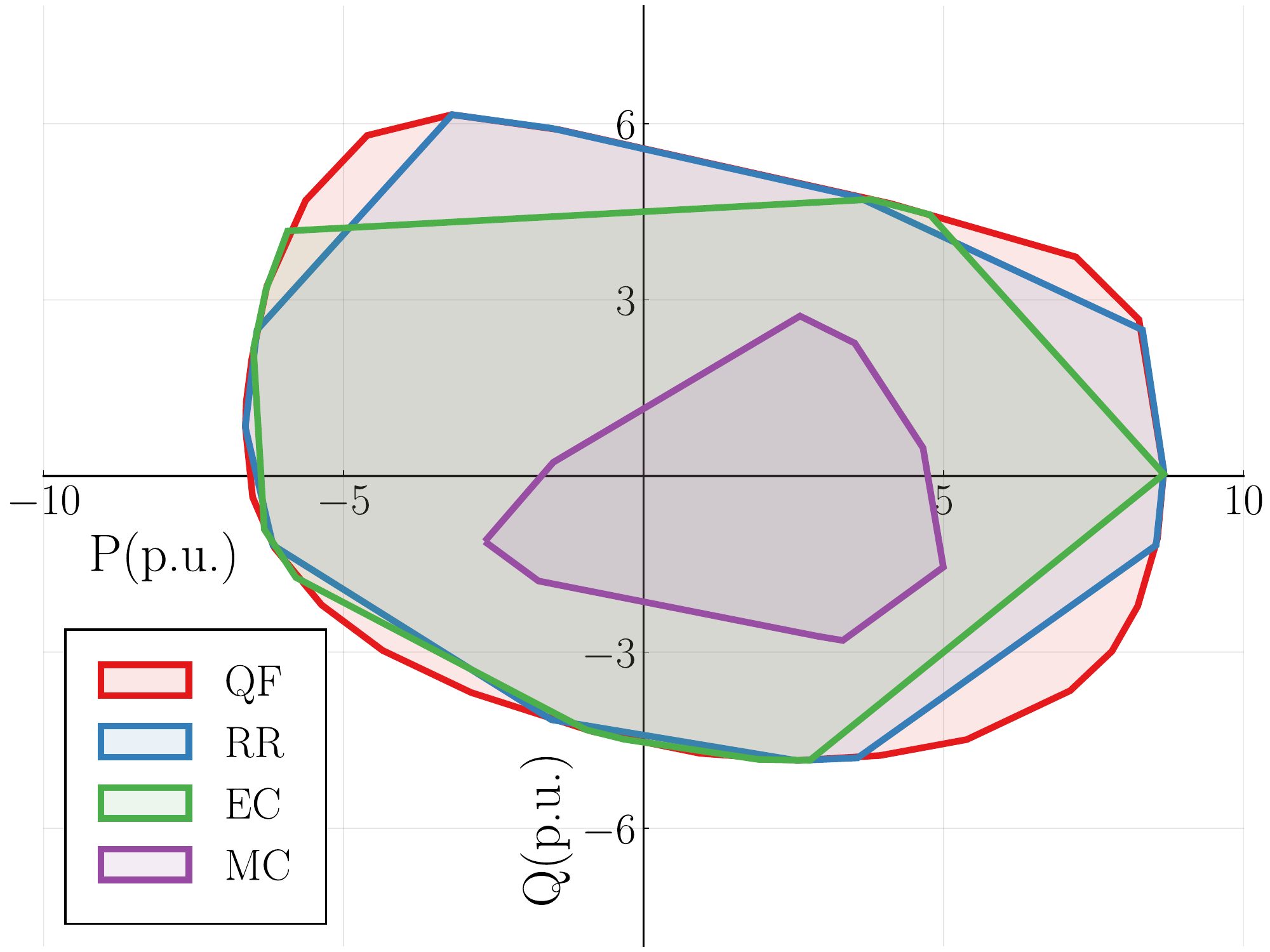}
    \caption{Comparison of the obtained flexibility regions by the proposed QuickFlex vs existing methods for the IEEE 123-node case study.}
    \label{fig:comp}
    \vspace{-0.5cm}
\end{figure}

\cref{tab:comp} summarizes the area of the TSO/DSO feasible region recovered by different methods.  Values reported are given as a relative area using the QuickFlex area as reference.
The \textbf{MC} method has the worst performance on computing the flexibility region for the three case studies, while the \textbf{RR} has the better  performance after the  \textbf{QF} (see \cref{fig:comp}). However, the  \textbf{EC} method is not competitive, recovering only 75\% of the true area in some cases for the same number of explored points. 
We want to highlight, once again, that the \textbf{QF} is the only method that can set an a priori error bound.

\section{Conclusions}\label{Sec:Conclusions}

We proposed the QuickFlex algorithm to construct the equivalent aggregated flexibility region of a distribution network at the TSO/DSO interface.  The QuickFlex algorithm has been proved to be effective in calculating the feasibility regions of medium-sized networks with a tolerance of less than $1\%$ with less than 10 points and tolerance of less than $0.1\%$ with less than 30 points. The SOC DistFlow relaxation and the LinDistFlow approximation used in modeling distribution network power flows greatly overestimate the feasible flexibility regions up to five and two times more than the true region, respectively. These overestimates can lead to misleading assumptions about the flexibility of the distribution system. We compared the QuickFlex algorithm with three existing methods in the literature, evaluating the area calculated by the methods in the same number of computations. The computational studies show that the QuickFlex method presented higher accuracy than the existing methods. 

\bibliographystyle{IEEEtran}
\bibliography{my_references}

\end{document}